# Synthesis and Thermal Stability of Cubic ZnO in the Salt Nanocomposites

P.S. Sokolov,[1,2] A.N. Baranov,[3] Zh.V. Dobrokhotova,[4] V.L. Solozhenko [1*]

[1] *LSPM-CNRS, Université Paris Nord, 93430 Villetaneuse, France*

[2] *Department of Materials Science, Moscow State University, 119991 Moscow, Russia*

[3] *Chemistry Department, Moscow State University, 119991 Moscow, Russia*

[4] *Kurnakov Institute of General and Inorganic Chemistry, RAS, 119991 Moscow, Russia*

**Abstract**

*Cubic zinc oxide (rs-ZnO), metastable under normal conditions was synthesized from the wurtzite modification (w-ZnO) at 7.7 GPa and ~800 K in the form of nanoparticles isolated in the NaCl matrix. The phase transition rs-ZnO → w-ZnO in nanocrystalline zinc oxide under ambient pressure was experimentally studied for the first time by differential scanning calorimetry and high-temperature X-ray diffraction. It was shown that the transition occurs in the 370-430 K temperature range and its enthalpy at 400 K is $-10.2 \pm 0.5$ kJ mol$^{-1}$.*

*\*e-mail: vladimir.solozhenko@univ-paris13.fr*

**Introduction**

Hexagonal (P6$_3$mc) zinc oxide with the wurtzite structure, *w*-ZnO, is a direct-band semiconductor ($E_g$ = 3.4 eV) with the highly ionic character of chemical bond [1], whereas cubic (Fm3m) zinc oxide with the rock salt structure, *rs*-ZnO, is an indirect semiconductor with the band gap energy $E_g$ ~ 2.7 eV at 11 GPa [2].

Under normal conditions, wurtzite ZnO is thermodynamically stable and at pressures of about 9 GPa and room temperature transforms into the cubic modification [3]. The reverse transition is observed only upon the pressure drop down to 2 GPa [4, 5], indicating considerable hysteresis between the direct (*w* → *rs*) and reverse (*rs* → *w*) phase transitions in ZnO at room temperature. The hysteresis width decreases with the temperature increase, and above 1200 K the branches of the direct and reverse transition are brought together under a pressure of about 6 GPa [4, 5], which can be considered the equilibrium pressure of this phase transition. The cubic phase of zinc oxide is stable only at pressures above 2 GPa and cannot be quenched down to ambient pressure [4, 5].

It was experimentally shown [6-8] that for nanocrystalline *w*-ZnO the direct transition occurs under higher (> 9 GPa) pressures and the samples with a particle size of ~12 nm treated at pressured higher than 15 GPa retain cubic structure after decompression [6, 7]. However, since all experiments mentioned were carried out in diamond anvil cells, it cannot be excluded that residual pressures are retained in the sample after pressure release.

The purpose of the present work is to study the possibility to stabilize cubic zinc oxide synthesized under relatively low (~7 GPa) pressures in the sodium chloride matrix.

**Experimental**

Sodium chloride (special-purity grade) and monodispersed nanoparticles of wurtzite zinc oxide with the average size ~50 nm, synthesized by the thermal decomposition of zinc acetate in diethylene glycol by the earlier described method [9], were used for the preparation of the starting mixtures. A mixture of *w*-ZnO and NaCl nanoparticles was ground in an agate mortar and then molded in a steel press mould under a pressure of 1250 bar.

Cubic ZnO was synthesized in a toroid-type high-pressure apparatus [10] at 7.7 GPa in the 700-900 K temperature range. The cell was calibrated against pressure using the Bi$_{III-IV}$ phase transition (7.7 GPa at room temperature), whereas temperature calibration was performed using Pt10%Rh—Pt thermocouple without correction for the pressure effect on the thermocouple emf.

Gold capsules were used to insulate the samples from the high-pressure assembly. The samples were gradually compressed at room temperature, and then the temperature was increased up to the required value. After isothermal heating for 15 min, the samples were quenched by switching off the power and slowly decompressed down to ambient pressure.

X-ray diffraction study of the synthesized samples and precise determination of the lattice parameters were performed on a TEXT 3000 INEL ($\lambda = 1.54056$ Å) diffractometer. A sample of $LaB_6$ ($a = 4.15695$ Å) was used as standard for detector adjustment. Lattice parameters were determined by the full-profile analysis with allowance for corrections to X-ray absorption and the shift of the sample surface from the diffraction plane. The size of ZnO crystallites was determined by the Debye—Scherrer equation taking into accounts the instrumental function.

The quenched samples were studied *in situ* at temperatures up to 1000 K *in vacuo* by X-ray diffraction on a Rigaku D/MAX 2500 diffractometer (Cu $K\alpha_{1,2}$ radiation) in the 30-65º (2Θ) range using an HT-1500 high-temperature attachment. The 2Θ scan rate was 5 K min$^{-1}$ at a linear heating rate of 2 K min$^{-1}$, which corresponded to the change in the sample temperature by 15 K during collection of each diffraction pattern.

The differential scanning calorimetry (DSC) study of the quenched samples in the 300-900 K temperature range was carried out using a Netzsch DSC 204 F1 calorimeter at continuous heating with rates of 2, 5 and 10 K min$^{-1}$ in high-purity (> 99.998%) argon. Aluminum ampules were used as sample holders. The calorimeter was calibrated by the phase transitions of the standard substances (Hg, $KNO_3$, In, Sn, Bi, and CsCl with purity not lower than 99.99%). The experimental data were processed using the NETZSCH Proteus Analysis program package.

**Results and discussion**

At 7.7 GPa and temperatures below 1000 K when pristine *w*-ZnO nanopowder was used as a starting material, the quenching resulted in the formation of a mixture of *rs*-ZnO and *w*-ZnO in various ratios rather than single-phase cubic zinc oxide. Therefore, all further experiments were carried out only with *w*-ZnO in the NaCl matrix.

Cubic ZnO was synthesized at 700—900 K which allowed us to ensure the completeness of the *w*-ZnO → *rs*-ZnO phase transition at 7.7 GPa, and to avoid an increase in the ZnO nanoparticle sizes at higher temperatures.

According to the X-ray diffraction data of the quenched samples, the complete stabilization of the cubic structure of zinc oxide nanoparticles in the sodium chloride matrix is observed only starting from the ZnO/NaCl ratio 1 : 3 (see Table 1). The diffraction pattern of the *rs*-ZnO/NaCl nanocomposite (1 : 3) with the maximum content of zinc oxide (25 wt. %) is presented in Fig. 1. All reflections in the diffraction patterns are assigned either to *rs*-ZnO (JCPDS No. 77-0191; Fm3m, $a$ = 4.28 Å) or to NaCl (JCPDS No. 05-0628; Fm3m, $a$ = 5.642 Å); no reflections of other phases, including *w*-ZnO, are observed. The average sizes of coherent scattering areas of *rs*-ZnO, estimated from the reflections broadenings in the diffraction patterns by the Debye-Scherrer equation, are almost the same for all salt nanocomposites listed in Table 1, being ~50 nm. This indicates that the sizes of the ZnO nanoparticles do not change during the phase transition at high pressures and temperatures.

When *w*-ZnO/NaCl initial mixture contains more than 33 wt. % zinc oxide, the quenched samples represent rs-ZnO and w-ZnO mixtures, *i.e.* only partial stabilization of the cubic structure of zinc oxide in the salt nanocomposites is observed after quenching. If *w*-ZnO micropowders (99.99%, Aldrich, 325 mesh) with an average particle size of ~44 μm are used as the starting material, the cubic ZnO phase can not be quenched down irrespective to the *w*-ZnO/NaCl ratios in the starting mixtures. This indicates that zinc oxide is characterized by some critical particle size, above which no stabilization of *rs*-ZnO can be achieved in the salt nanocomposites under normal conditions. The results obtained allow the conclusion that the salt matrix plays the key role in the stabilization of the *rs*-ZnO nanoparticles synthesized at high pressures and temperatures after their quenching.

In a special series of experiments we have shown that the dissolution of the salt matrix of the *rs*-ZnO/NaCl nanocomposites in distilled water initiates the reverse phase transition *rs*-ZnO → *w*-ZnO, and after the complete removal of NaCl the sample represents *w*-ZnO with trace amounts of *rs*-ZnO.

At room temperature the *rs*-ZnO nanoparticles in the sodium chloride matrix exhibit no tendency to the transition to *w*-ZnO for at least 5-6 months. For a longer storage period (1 year), the partial reverse transition *rs*-ZnO → *w*-ZnO is observed in the samples, and finally, after two years, the nanoparticles of cubic zinc oxide are completely transformed into wurtzite zinc oxide. Thus, it can be asserted that at room temperature the salt matrix kinetically hinders the reverse phase transition of zinc oxide from the metastable cubic modification to the thermodynamically stable wurtzite one.

The thermal stability of the as-synthesized *rs*-ZnO/NaCl nanocomposites of different composition (see Table 1) at ambient pressure was studied *in situ* by X-ray diffraction and differential scanning calorimetry. The diffraction patterns of the *rs*-ZnO/NaCl nanocomposite 1 : 3 collected *in situ* are presented in Fig. 2. It can be seen in the left part of Fig. 2 that the (*002*) reflection of *w*-ZnO appears at 365 K and its intensity increases during further temperature increase. The right part of Fig. 2 shows the corresponding decrease in the intensity of the (*200*) reflection of *rs*-ZnO, which disappears completely at 425 K. Thus, two phases of zinc oxide in the NaCl matrix coexist in the 365–425 K temperature range.

Fig. 3 shows the temperature dependence of the integral intensity of (*200*) reflection of *rs*-ZnO normalized to the corresponding 300 K value. The curve has the classical S-like shape, *i.e.*, at low temperature the transition rate is low, then it increases sharply with temperature, passes through a maximum at ~384 K, and further decreases down to zero at temperatures above 430 K.

Upon further heating of the nanocomposite in the temperature range from 430 to 970 K, the intensities of *w*-ZnO reflection increase noticeably, most likely, due to an increase in degree of crystallinity of the phase. However, according to the DSC data, this process is not accompanied by any thermal effect.

The results of the thermoanalytical study of the 1 : 3 and 1 : 5 salt nanocomposites are presented in Fig. 4 and Table 1. At temperatures above 380—430 K, the DSC curves contain the pronounced exothermic effect, which can unambiguously ascribed to the irreversible structural transition of cubic zinc oxide into the wurtzite phase, which is in good agreement with the results of high-temperature *in situ* X-ray diffraction study. As can be seen from the Table 1, for all *rs*-ZnO/NaCl nanocomposites the increase in the heating rate is accompanied by an increase in the peak temperature.

Along with the main exothermic effect, the thermoanalytical curve of the nanocomposite with the maximum content of *rs*-ZnO (1 : 3) contains the weak exothermic effect at 350-370 K (see Fig. 4, *b*), which can be attributed to the phase transition of submicronic aggregates of the *rs*-ZnO nanoparticles. It can be assumed that the *rs*-ZnO → *w*-ZnO transition in such aggregates occurs at lower temperatures than in the case of rs-ZnO nanoparticles isolated in the salt matrix. For a more "dilute" composition 1 : 4 , this effect is less pronounced and is completely absent for the composition with *rs*-ZnO/NaCl ratio 1 : 5 (see Fig. 4, a). No other processes accompanied by the heat release or absorption were observed up to 870 K. Thus, based on the data of DSC and high-temperature X-ray diffraction, we can conclude that the structural transition Fm3m → P6$_3$mc observed for zinc oxide in the sodium chloride matrix at ambient pressure is the monotropic first-order phase transition [11] that occurs without formation of any intermediate phase.

For the 1 : 3 and 1 : 4 *rs*-ZnO/NaCl nanocompositions, the thermal effect of the phase transition at a given heating rate was determined as the sum of two observed exothermic effects taking place at 350-370 K and 380—430 K, respectively. The thermal effect values obtained at different heating rates were averaged for each composition (see Table 1). The enthalpy of the phase transition at 400 K was determined by averaging over all experimental values of thermal effects and is equal to -10.2±0.5 kJ mol$^{-1}$.

The enthalpy of the reverse *rs*-ZnO → *w*-ZnO phase transition determined by us differs by an order of magnitude from the enthalpy values of the direct *w*-ZnO → *rs*-ZnO phase transition estimated from the data on ZnO solubility in borate melts ($\Delta_{tr}H°_{297\,K}$ = 24.5±3.6 kJ mol) [12, 13] and temperature dependences of the emf of the electrochemical cells with ZnO-based electrodes ($\Delta_{tr}H°_{1173\,K}$ = 17.4±1.3 kJ mol$^{-1}$) [14]; as well as from the $\Delta_{tr}H°_{297\,K}$ = 3.3 kJ mol$^{-1}$ value [3], calculated by the Clapeyron-Clausius equation from the experimental data on high-pressures synthesis of *rs*-ZnO. Unlike the direct calorimetric determination of the enthalpy of the *rs*-ZnO → *w*-ZnO phase transition performed in the present work, the enthalpy values reported earlier [3, 12-14] were obtained from indirect thermodynamic data and, hence, should be considered as estimates.

**Conclusions**

Thus, in the present work we have shown for the first time that the cubic phase of zinc oxide can be synthesized in the form of salt nanocomposites, which are kinetically stable under ambient pressure up to temperatures of about 370 K. This provides new possibilities for studying the physical and chemical properties of cubic ZnO. According to the DSC data, at ambient pressure the *rs*-ZnO → *w*-ZnO phase transition in the salt nanocomposites occurs in the 370-430 K range, and its enthalpy at 400 K is –10.2±0.5 kJ mol$^{-1}$.

**Acknowledgements**

We thank O.O. Kurakevych for help in high-pressure experiments and V.A. Mukhanov for valuable comments. P.S. Sokolov is grateful to the French Government for financial support (Bourse de co-tutelle no. 1572-2007).

This work was financially supported by the Russian Foundation for Basic Research (Project No. 09-03-90442-Ukr_f_a).

**Table 1.** Results of the thermoanalytical study of the *rs*-ZnO/NaCl nanocomposites

| ZnO/NaCl [a] | Q (kJ mole$^{-1}$) [b] | $T_{max}$ (K) at $v_T$ [c] | | |
|:---:|:---:|:---:|:---:|:---:|
| | | 2 | 5 | 10 |
| 1 : 5 | -10.2 ± 0.4 | 384 | 392 | 399 |
| 1 : 4 | -10.9 ± 0.4 | 387 | — | 404 |
| 1 : 3 | -9.8 ± 0.4 | 390 | 399 | 406 |

[a] Weight ratio

[b] Q is the thermal effect

[c] $T_{max}$ is the peak temperature of the main thermal effect at different heating rates, $v_T$ / K min$^{-1}$.

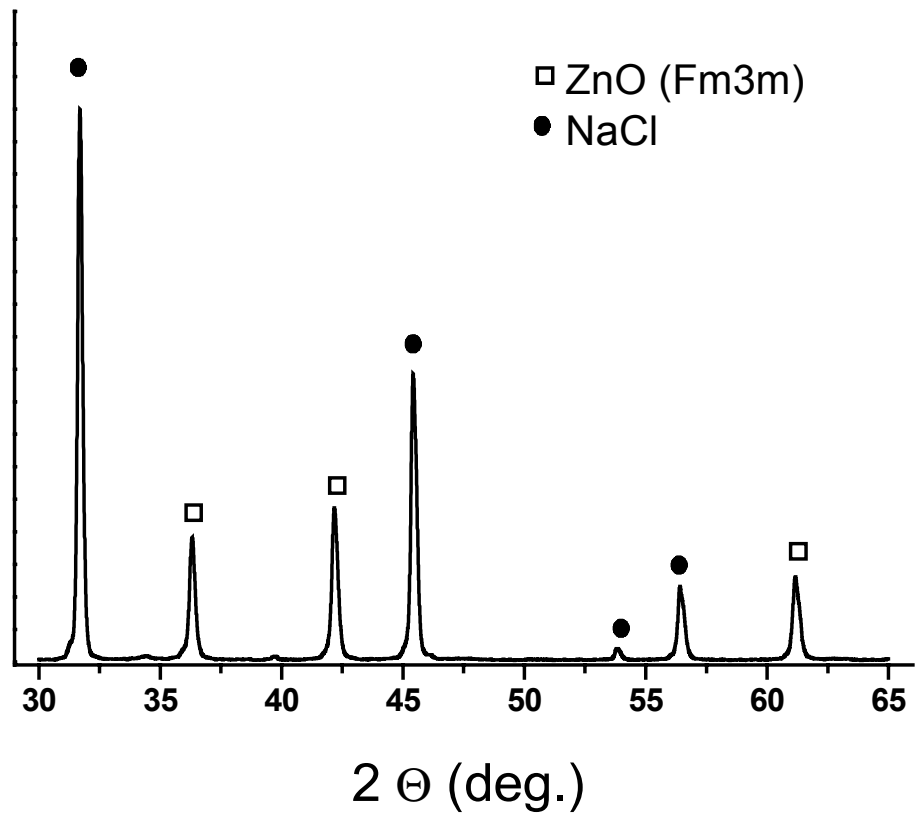

**Fig. 1.** Diffraction pattern of the *rs*-ZnO/NaCl nanocomposite (1 : 3) synthesized at 7.7 GPa and 800 K.

**Fig. 2.** Regions of the diffraction patterns of the *rs*-ZnO/NaCl nanocomposite (1 : 3) taken in situ in the course of linear heating with a rate of 2 K min$^{-1}$. Note: The reflections (*200*) of *rs*-ZnO and (*002*) of *w*-ZnO were chosen as the most intense non-overlapping reflections.

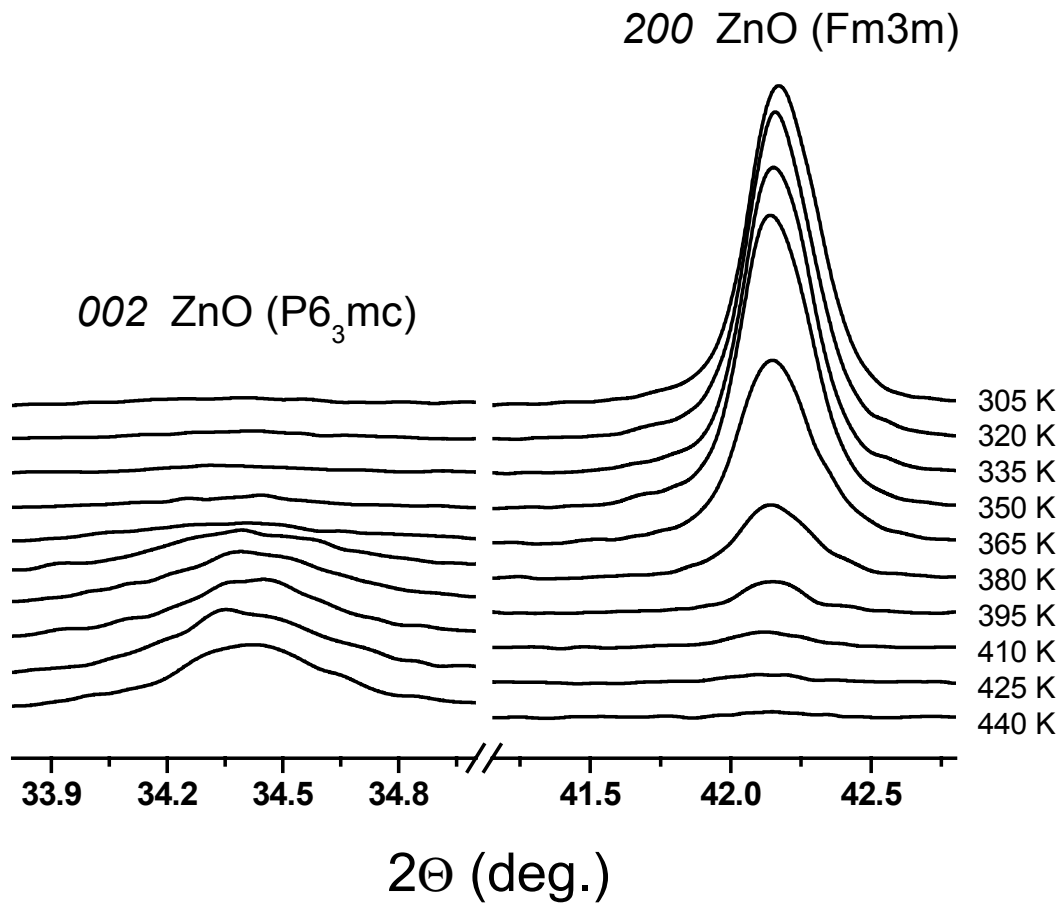

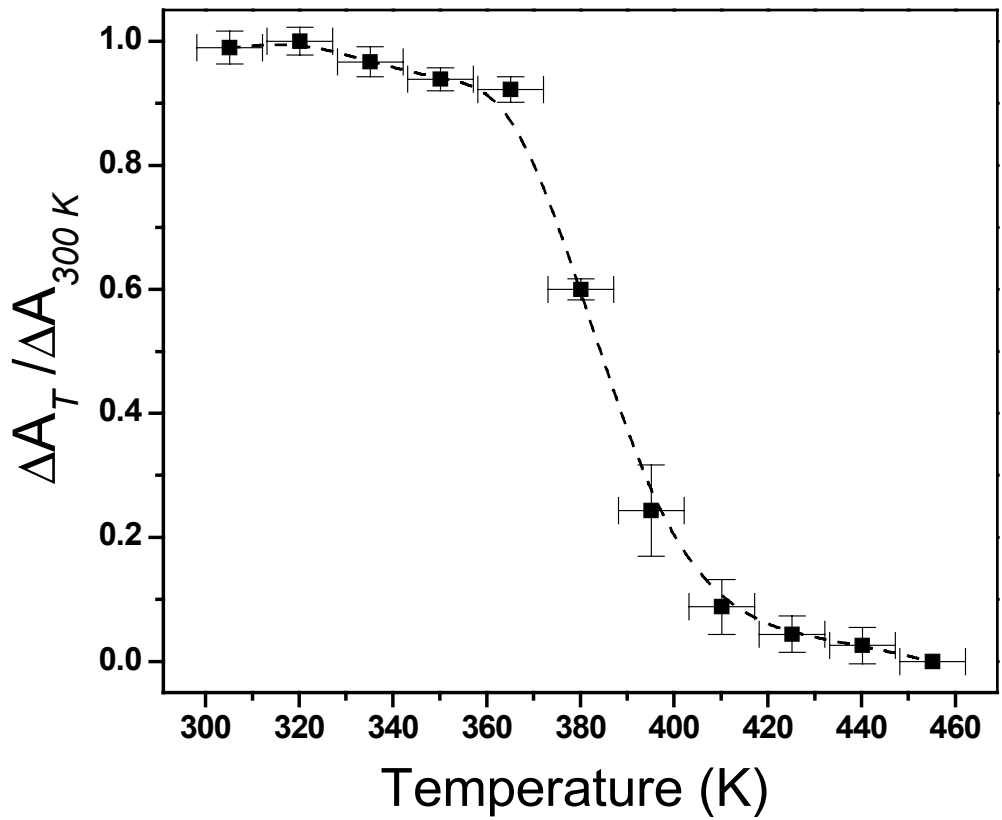

**Fig. 3.** Temperature dependence of the integral intensity of (*200*) reflection of *rs*-ZnO normalized to the corresponding 300 K value.

**Fig. 4.** DSC curves of the *rs*-ZnO/NaCl nanocomposites of the 1 : 5 (*a*) and 1 : 3 (*b*) composition recorded at the heating rate 2 (*1*), 5 (*2*), and 10 (*3*) K min$^{-1}$ ($Q$ is the specific heat flux).

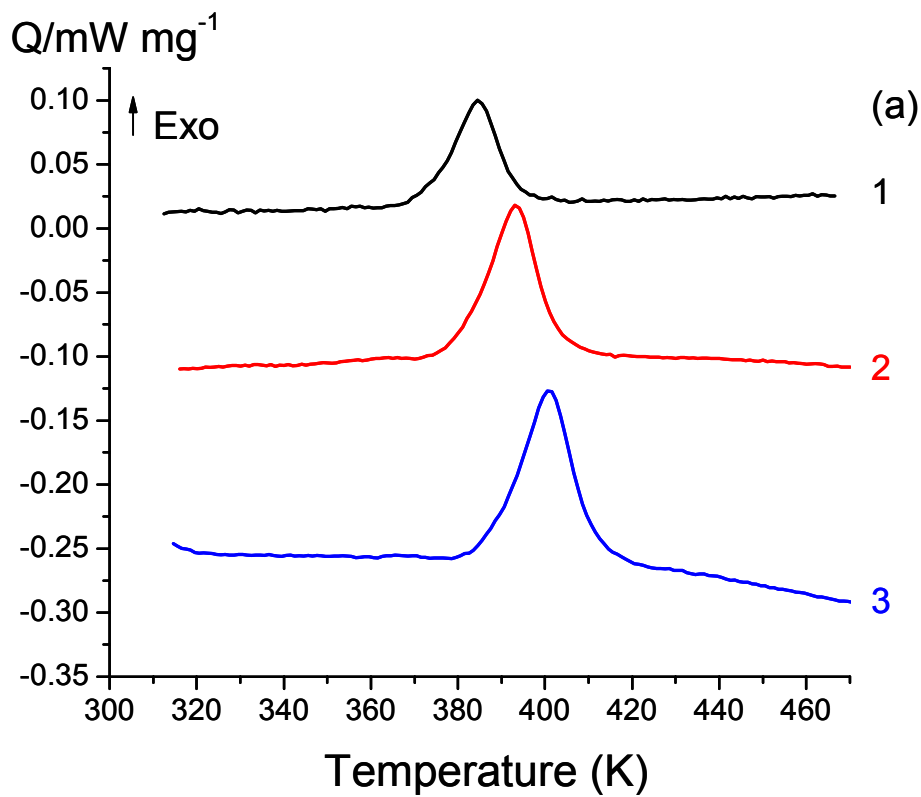

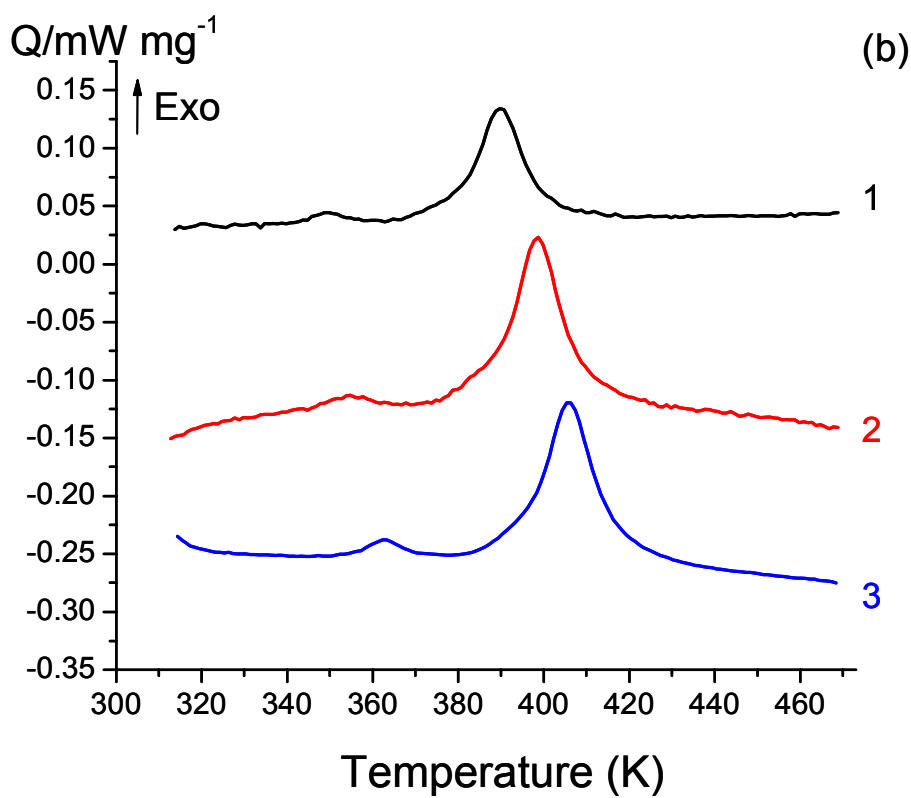